\documentclass[apJ, twocolumn]{aastex631}
\usepackage{graphicx}
\usepackage{amsmath}
\usepackage{color}
\usepackage{comment}
\usepackage{units}
\usepackage[normalem]{ulem}
\usepackage{hyperref}

\newcommand{\dynesty}{\textsc{dynesty}~}
\newcommand{\bilby}{\textsc{bilby}~}
\newcommand{\imrphenompv}{\textsc{IMRPhenomPv2}~}

\newcommand{\SPA}{School of Physics and Astronomy, Monash University, Clayton VIC 3800, Australia}
\newcommand{\OzGravMonash}{OzGrav: The ARC Centre of Excellence for Gravitational Wave Discovery, Clayton VIC 3800, Australia}

\begin{document}

\title{A search for extra polarisations using a Gaussian process in Gravitational-Wave Transient Catalogue 3}

\author{Shun Yin Cheung}
\email{shun.cheung@monash.edu}
\affiliation{\SPA}
\affiliation{\OzGravMonash}

\author{Lachlan Passenger}
\email{lachlan.passenger@monash.edu}
\affiliation{\SPA}
\affiliation{\OzGravMonash}

\author{Paul D. Lasky}
\affiliation{\SPA}
\affiliation{\OzGravMonash}

\author{Eric Thrane}
\affiliation{\SPA}
\affiliation{\OzGravMonash}

\begin{abstract}

General relativity predicts that gravitational waves are described by two polarisation states: the plus $+$ state and cross $\times$ state.
However, alternate theories of gravity allow up to six polarisations. We employ the gravitational-wave \textit{null stream}, a linear combination of three or more detectors where the $+$ and $\times$ signals add to zero, leaving behind noise and---potentially---gravitational waves in non-standard polarisation states. We develop a Gaussian process model to search for extra polarisations beyond general relativity. Using data from 42 three-detector events from LIGO--Virgo--KAGRA's Third Gravitational-Wave Transient Catalogue, we find no evidence of non-standard polarisations. We set upper limits on the fractional deviation in gravitational-wave strain to be as low as 0.39 at 90\% credibility for the event GW190602\_175927.

\end{abstract}


\section{Introduction}
The first detection of gravitational waves in 2015 by the LIGO--Virgo--KAGRA (LVK) collaboration has provided a new way to test general relativity (GR) in the strong-field regime \citep{Abbott2016}.
General relativity predicts only two gravitational-wave polarisations: plus $+$ and cross $\times$. However, the most general metric theories of gravity contain up to six polarisations: scalar breathing $b$ and longitudinal $L$ polarisations, and vector $x$ and $y$ polarisations---in addition to the usual tensor plus and cross polarisations.
Different alternative theories of gravity contain different sets of polarisations. 

Examples of alternate theories include Brans-Dicke theory, which predicts an extra breathing polarisation \citep{brans_machs_1961}, Einstein-aether theory predicting five polarisations \citep{jacobson_einstein-aether_2004} and TeVeS that predicts all six polarisations \citep{bekenstein_relativistic_2005}.
A detection of scalar and/or vector polarisations would imply physics beyond GR \citep{Eardley_TGR,will_confrontation_2014}. 

Different methods have been developed to search for extra polarisations. One common technique is to construct a linear combination of three or more strain time series that removes the tensor polarisations while preserving some signal from the extra polarisations. This data stream is referred to as a \emph{null stream} \citep{gursel_near_1989}.
Previous work has used excess power algorithms to search for non-standard polarisations in the null streams of events in the Third Gravitational-Wave Transient Catalogue (GWTC-3) \citep{wong_null-stream-based_2021, TGR_GWTC2, TGR_GWTC3, GWTC-3_paper}. 
Other literature proposes to use the null stream in next-generation ground-based and space-based observatories to search for extra polarisations \citep{wang_observing_2021, hu_probing_2023}.

Tests of GR can be categorised as either \textit{prescriptive} or \textit{phenomenological}. Prescriptive models make predictions for the form of the deviations based on a theoretical framework. This includes tests of PN coefficients in the inspiral phase \citep{Mishra_2010, Li_2012, TGR_GWTC1, TGR_GWTC2, TGR_GWTC3}, tests of gravitational-wave propagation \citep{Saeed_Lorentz_violating} and tests of post-Einsteinian inspiral phases \citep{Yunes_2009,cornish_gravitational_2011, chatziioannou_model-independent_2012}.
With many alternative theories and little compelling evidence for any specific theory \citep{yunes_gravitational-wave_2013}, it is difficult to define a prescriptive model to find a deviation. 
Phenomenological models make less specific predictions about the signal. Examples include consistency tests and searches for excess power in the null stream \citep[e.g.,][]{TGR_GWTC2, TGR_GWTC3}. 

Gaussian processes provide flexibility to model different possible deviations by assuming the data is drawn from a multivariate Gaussian probability distribution (see \citet{Aigrain_2023} for a recent review).
The Gaussian process is described by a covariance matrix, which can be written in terms of a kernel function, which can be chosen to incorporate prior beliefs about the signal.
Gaussian processes have been used in gravitational-wave astronomy to model glitches \citep{GP_glitch} and to account for uncertainties in waveform approximants \citep{moore_2016, liu_gp_error}.

In this paper, we use a recently-developed Gaussian process to model deviations of GR \citep{Passenger_2025}. 
In our model, the kernel enforces the following prior beliefs: the signal is likely localised near the gravitational-wave merger; it has some characteristic frequency; and it is not necessarily symmetric about the merger.
We analyse the null stream of GWTC-3 events using a Gaussian process model to search for extra polarisations. 

Our paper is organised as follows. We lay out our method in Section~\ref{formalism}. We verify that our method is working as intended with several simulations in Section~\ref{sec:simulations}. We present and discuss our results on the GWTC-3 data in Section~\ref{sec:data}. We provide concluding remarks and discuss possibilities for future work in Section~\ref{sec:conclusion}. 

\section{Formalism}\label{formalism}
\subsection{The null stream}
If GR is a complete description of gravity, then the null stream is a signal-free data stream, constructed by a linear combination of multiple detectors' strain series \citep{gursel_near_1989}. In a network with $N_\text{det}$ non-aligned detectors and $N_\text{pol}$ polarisations, we can create ($N_\text{det}-N_\text{pol}$) null streams. Hence, we require a minimum of three detectors to null the two tensor polarisations from GR. 

The null stream depends on four extrinsic parameters, which affect how the $+$ and $\times$ polarisations couple to the interferometers: $\theta=\{\alpha, \delta, \psi, t_c\}$.
Here, $\alpha$ is the right ascension, $\delta$ is the declination, $\psi$ is the polarisation angle, and $t_c$ is the time of coalescence.

A null stream of the tensor polarisations is constructed using 
\begin{align}\label{eq:null2}
    d_\text{null}(t|\theta) 
    = & d_1(t)-\eta(\theta)d_2(t+\tau_{12})-\zeta(\theta)d_3(t+\tau_{13}),
\end{align}
where $\tau_{ij}$ is the time delay between detectors $i$ and $j$, which depends implicitly on $\alpha, \delta, t_c$.
Meanwhile, $\eta$ and $\zeta$ are coefficients that depend on the gravitational-wave antenna response functions $F_{+, \times}(\theta)$ for the three observatories~\citep{gursel_near_1989}:
\begin{align}\label{eq:eta}
    \eta(\theta)=-\frac{F^3_{\times} F^1_{+} - F^3_+ F^1_\times}{F^2_\times F^3_+ - F^2_+ F^3_\times},\\ 
    \zeta(\theta)=-\frac{F^1_{\times} F^2_{+} - F^1_+ F^2_\times}{F^2_\times F^3_+ - F^2_+ F^3_\times},
\end{align}
where the superscript on the antenna function $F_{+, \times}$ indicate the detector. 
The $F$ factors depend implicitly on $\alpha, \delta, \psi$.

If additional polarisations exist in our data, our null stream can be expressed as 
\begin{equation}\label{eq:null_stream}
    d_\text{null} = n_\text{null}+\delta s.
\end{equation}
where $n_{\text{null}}$ is the noise in the null stream, which is characterised by the noise power spectral density
\begin{equation}\label{eq:PSD}
    P_\text{null}(\theta, f)= P_1(f)+\eta(\theta)^2 P_2(f) + \zeta(\theta)^2 P_3(f),
\end{equation}
where $P_i(f)$ is the noise power spectral density of the $i^\text{th}$ detector. In Equation~\ref{eq:null_stream}, 
$\delta s$ is the signal for non-standard polarisations observed in the null stream, which can be expressed as

\begin{align}\label{eq:delta_s}
    \delta s = \sum_m \left(F^1_m - \eta F^2_m - \zeta F^3_m\right) h_m,
\end{align} where $m = \left\{ b, L, x, y\right\}$ is the set of all non-standard polarisations, $h_m$ are the strain from different polarisations, and the $F^i_m$ are the antenna response functions to the different polarisations in the $i^\text{th}$ detector.

\subsection{Gaussian process}

\subsubsection{Signal model}

We model the extra polarisations with a Gaussian process so that the signal is described by a covariance matrix:
\begin{align}\label{eq:S}
    \textbf{\textit{S}}_{ij} \equiv \langle \delta s^*(f_i) \, \delta s(f_j) \rangle .
\end{align}
We assume that each polarisation is uncorrelated with each other so that
\begin{align}
    \langle h_m^*(f_i) \, h_{n}(f_j) \rangle 
    = \delta_{mn} \textbf{\textit{K}}_{ij},
\end{align}
where $m$ and $n$ label the non-standard polarisations, $\delta$ is the Kronecker delta and $\textbf{\textit{K}}$ is the kernel; we describe the kernel design in the next section. The signal covariance matrix is determined by the kernel multiplied by a combination of the antenna functions of the non-standard polarisations, 

\begin{align}\label{eq:Sij}
    \textbf{\textit{S}}_{ij} = & 
    \left( \sum_{m} (F_m^1 - \eta F_m^2 - \zeta F_m^3)^2 
    \right) \textbf{\textit{K}}_{ij}.
\end{align}

\subsubsection{Kernel design}
Our kernel, proposed in \citet{Passenger_2025}, is designed to achieve the following goals:
\begin{itemize}
    \item Deviations from GR should be localised in time so that $\delta s (t)$ is maximal near the merger time where one expects the largest deviations from GR, and goes to zero on some timescale before and after the merger.
    \item The deviation has some characteristic frequency, which is similar to the merger frequency.
    \item It is not necessary that $\delta s (t)$ be symmetric in time.
\end{itemize}
Following \citet{Passenger_2025}, we use the following kernel:
\begin{align}\label{eq:Kt}
    \textbf{\textit{K}}_{ij} = & k(t_i, t_j) \nonumber\\
    = &
    k_0 e^{-f_0^2 (t_i^2+t_j^2) / 2 w^2}
    \cos\left(2\pi f_0 \tau_{ij}\right)
    e^{-f_0^2 \tau_{ij}^2 / 2l^2} ,
\end{align}
where
\begin{align}
    \tau_{ij} \equiv | t_i - t_j | ,
\end{align}
is the absolute value of the difference of two sample times $t_i \text{ and } t_j$.

\begin{figure}[h]
    \centering
    \includegraphics[width=0.85\linewidth]{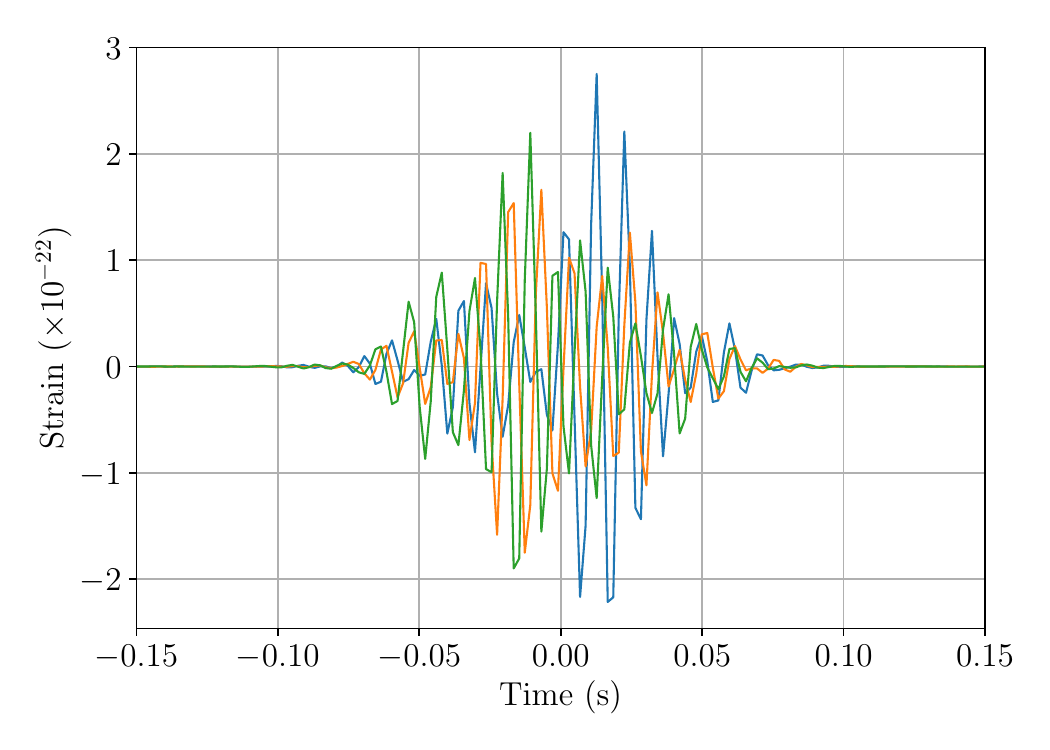}
    \caption{Each coloured curve is a strain time series randomly drawn from an example kernel with parameters $k_0=1\times 10^{-44}$, $w=3$, $f_0=\unit[100]{Hz}$ and $l=1$. The kernel enforces our prior belief that deviations from GR are most likely to occur near the merger with some characteristic frequency similar to the frequency at merger.}
    \label{fig:kernel_example}
\end{figure}

The kernel is characterised by several parameters:

\begin{itemize}
    \item The scale factor $k_0$ controls the amplitude of $\delta s (t)$.
    \item The width $w$ sets the duration of $\delta s (t)$.
    \item The characteristic frequency $f_0$ determines the oscillation time scale of $\delta s (t)$.
    \item The coherence length $l$ determines how many cycles the frequency of $\delta s (t)$ remains coherent. Larger values of $l$ result in more sinusoidal waveforms whereas smaller values result in more stochastic waveforms.
\end{itemize}
Figure \ref{fig:kernel_example} shows multiple draws from the combined kernel. Each draw (coloured curve) in Fig.~\ref{fig:kernel_example} shows the signal being localised near the merger, periodic and asymmetric. 

\subsection{Likelihood function}
We employ the likelihood function from \cite{Passenger_2025}
\begin{align}\label{eq:L}
    {\cal L}(d_{\text{null}} | \Lambda) = &
    \frac{1}{2\pi \det \textbf{\textit{C}}(\Lambda)} \exp 
    \left(-\left|\frac{1}{2} d_{\text{null}}^\dagger \textbf{\textit{C}}^{-1} (\Lambda) d_{\text{null}} \right|\right),
\end{align}
where $\textbf{\textit{C}}(\Lambda)$ is the total covariance matrix
\begin{align}\label{eq:C}
    \textbf{\textit{C}}(\Lambda) = \textbf{\textit{S}}(\Lambda) + \textbf{\textit{N}} ,
\end{align}
and $\Lambda=\{k_0, w, f_0, l\}$ is the kernel parameters. The noise matrix $\textbf{\textit{N}}_{ij}$ is defined as

\begin{align}\label{eq:N}
    \textbf{\textit{N}}_{ij} \equiv & \langle n(f_i)^* \, n(f_j) \rangle \\
    = & \frac{1}{4 \Delta f} P_\text{null}(f_j) \, 
    \delta_{ij}.
\end{align}
Here, $\Delta f$ is the frequency resolution.

\subsection{Bayesian inference}

For all analyses in this paper, we perform parameter estimation on the null stream to obtain posterior samples for $\Lambda$. Each segment of null stream data is $\unit[2]{s}$ in duration, corresponding to $\pm \unit[1]{s}$ around the time of coalescence. If $k_0>0$ with high credibility, we take that as a sign of a deviation from GR.
We use the \dynesty nested sampler \citep{dynesty} that is included in the gravitational-wave Bayesian inference package \bilby \citep{Ashton_2019,bilby_gwtc1}. We use 1000 live points with a stopping condition of $\text{d}\log z < 0.1$.

\subsection{Marginalising over the extrinsic parameters}\label{sec:marginalisation_extrinsic}

The construction of the null stream is dependent on the sky location, polarisation angle and time of coalescence, which we infer from parameter estimation of the original binary merger signal. 
It follows that the data constructed from the maximum-likelihood estimator is only an approximate null stream.
In Appendix \ref{appendix_A}, we study the systematic error from this approximation.
We show that it is possible to marginalize over uncertainty in sky location, polarisation angle, and time of coalescence.
However, we argue that the maximum likelihood approximation is adequate for our present analysis.

\section{Simulations}\label{sec:simulations}

In order to validate the pipeline, we analyse simulated data. We simulate a signal (with only $+/\times$ polarisations) from a $\unit[30]{\text{M}_\odot}$ equal-mass, non-spinning binary system with a luminosity distance of \unit[400]{Mpc}. We construct our null stream assuming data from the LIGO Hanford, LIGO Livingston \citep{collaboration_advanced_2015} and Virgo observatories \citep{acernese_advanced_2014} with Gaussian noise at O4 design sensitivity \citep{noise_curve}. Furthermore, we assume the sky location is known perfectly, with right ascension and declination $(\alpha, \delta)=(8^\text{h}1^\text{m}, -74.5^\circ)$,  and time of coalescence is the same as GW150914.

\subsection{Breathing mode injection into detectors}\label{sec:breathing_mode_injection}
To validate our pipeline, we check if we can detect a signal in the null stream due to a
non-standard polarisation signal in the individual detectors. We inject a binary black hole signal with a GR-violating merger into all three detectors and construct a null stream. For the deviation of GR, we create a signal covariance matrix using the following kernel parameters, $k_0 = 8\times10^{-43}$, $f_0=\unit[128]{Hz}$, $w = 2 $ and $l=2$.
Then we draw $\delta s$ from the signal covariance matrix. The optimal signal-to-noise ratio (SNR) of the deviation in the null stream is 6.2. 

\begin{figure*}[ht]
    \centering
    \includegraphics[width=0.65\linewidth]{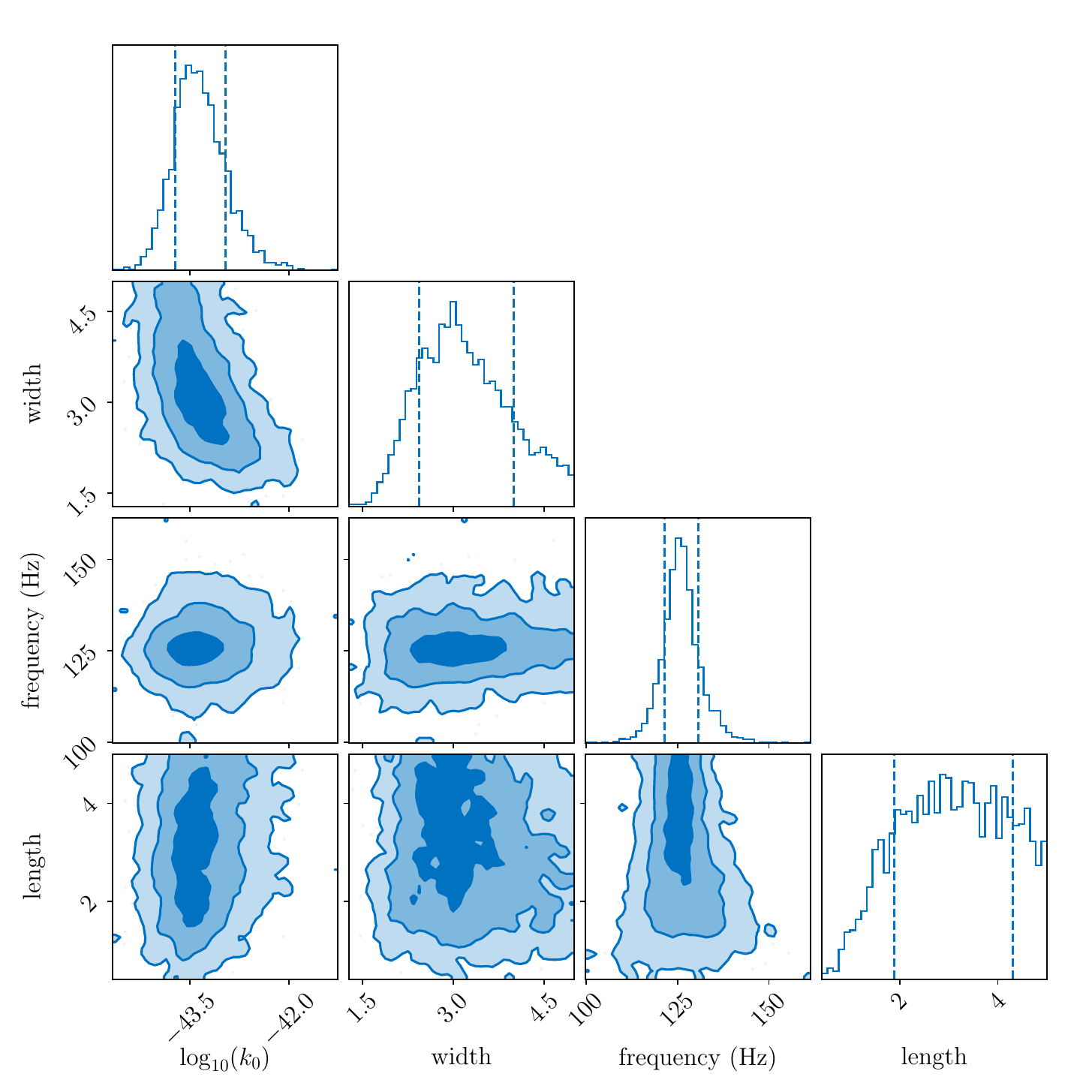}
    \caption{Posteriors of the kernel parameters for the breathing mode injection. We inject a draw from the kernel as a breathing mode signal into the individual detectors and analyse the resulting signal in the null stream. The $k_0$ posterior rules out the lower bound of the prior, indicating the Gaussian process is detecting a signal.}
    \label{fig:breathing_only_corner}
\end{figure*}

We take the data frequency band to be $\unit[50 - 512]{Hz}$. The lower limit is chosen to minimise the effect of lower frequency non-Gaussian noise, as \citet{Cheung_2024} found that it can complicate tests of GR. The upper limit is chosen as we do not expect the frequency of the deviation to be twice the ringdown frequency. An example is GW150914 with a ringdown frequency of $\sim \unit[260]{Hz}$, hence the upper bound is $\unit[520]{Hz}$, which we round to the nearest power of two, $\unit[512]{Hz}$. In our analysis of real data presented below, we do not analyse events with ringdown frequency greater than $\unit[512]{Hz}$.

We employ a log-uniform prior for $k_0$ between $10^{-47}$ and $10^{-41}$. The parameter $k_0$ determines the typical (square of the) strain amplitude. For the characteristic frequency $f_0$, we employ a uniform prior on the interval $\unit[50-1024]{Hz}$. The lower limit of this prior is the minimum frequency band of our data. The upper limit is double the maximum frequency band as we do not expect the frequency of the deviation to be much greater than twice our maximum ringdown frequency of $\unit[512]{Hz}$. The prior for the width parameter $w$ is uniform on the interval $(0.001, 5)$. The prior on the coherence length $l$ is uniform on the interval $(0.05, 5)$.

We carry out Bayesian inference to obtain posterior samples. Figure \ref{fig:breathing_only_corner} shows the posteriors of the kernel parameters. These show we rule out the lower bound of our $k_0$ prior at 90\% credibility. Next, we reconstruct the signal $\delta s$ by marginalising over the Gaussian process parameters using the equations

\begin{align}
    p( \delta s | \delta h)
    \propto & \sum_k
    \exp \Big( - \frac{1}{2}(\delta s-\mu_k)^\dagger \Sigma_k (\delta s-\mu_k) \Big),
\end{align}
which is a multivariate Gaussian distribution with mean $\mu$ and variance $\Sigma$
\begin{align}
    \mu_k &= (\textbf{\textit{N}}^{-1} + \textbf{\textit{S}}^{-1}(\Lambda_k))^{-1}\textbf{\textit{N}}^{-1}d_{\text{null}} ,\\
    \Sigma_k &= \textbf{\textit{N}}^{-1}+\textbf{\textit{S}}^{-1}(\Lambda_k).
\end{align}
For additional details, see \citet{Passenger_2025}.

\begin{figure}[h]
    \centering
    \includegraphics[width=0.95\linewidth]{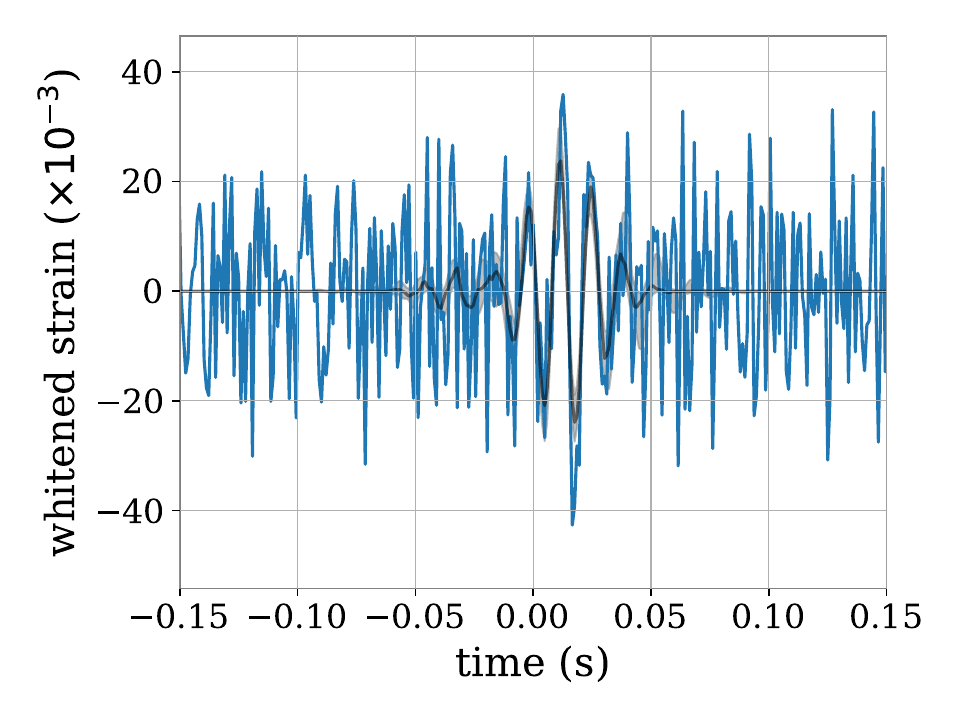}
    \caption{A reconstruction of the deviation signal $\delta s$ (black) for a breathing polarisation injection into individual detectors plotted along with the whitened data (blue). The data is within the 90\% credible interval of $\delta s$ (grey shaded region) estimated by the Gaussian process, showing the Gaussian process can capture the signal in the data.}
    \label{fig:breathing_only_reconstruction}
\end{figure}

The reconstructed waveform, shown in Figure \ref{fig:breathing_only_reconstruction}, is inconsistent with no signal (a zero line) within a 90\% credible interval. 

We quantify the statistical significance of deviations from GR with a  Bayes factor:
\begin{equation}\label{eq:BF}
    \mathcal{B}= \frac{\int dk_0 \, \mathcal{L}(d|k_0)\pi(k_0)}{\mathcal{L}(d|k_0=0)} .
\end{equation}
The numerator is the Bayesian evidence for GR violation ($k_0>0$) while the denominator is the Bayesian evidence for GR ($k_0=0$).
Here, $\pi(k_0)$ is the prior on $k_0$. A natural log Bayes factor of $\ln\mathcal{B}= 8$ is sometimes used as a threshold for when one model is strongly preferred over another \citep[e.g.][]{intro}. For this injection, we calculate a natural log Bayes factor of $\ln\mathcal{B}=8.8$. Hence, we conclude our Gaussian process can find a signal in the null stream if a non-standard polarisation is present in the individual detectors.

\subsection{Testing for ``safety''}
\begin{figure}
    \centering
    \includegraphics[width=0.9\linewidth]{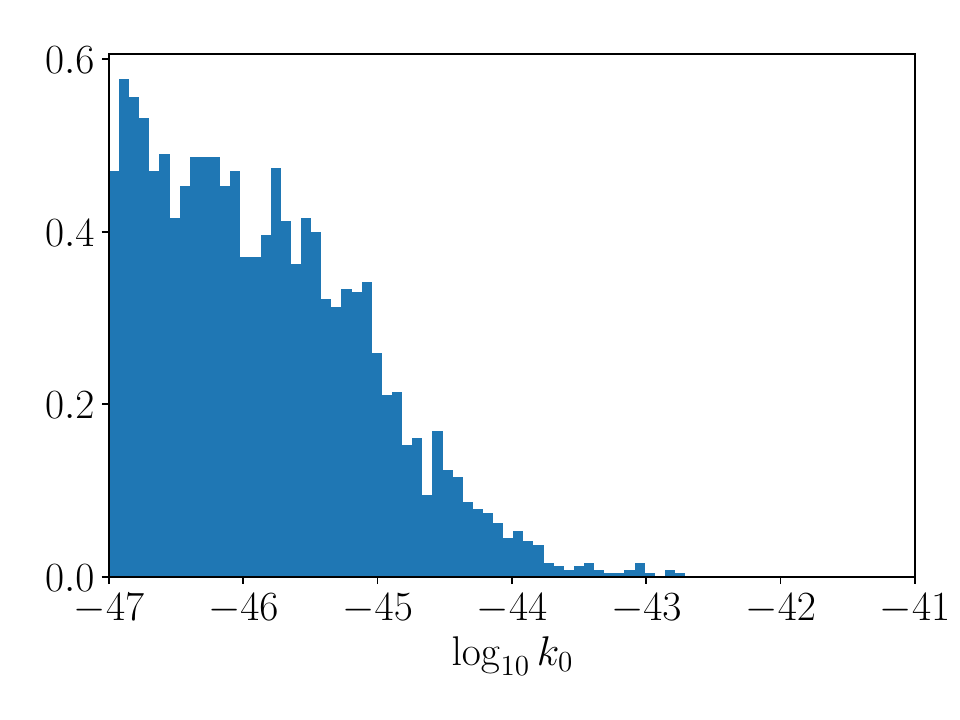}
    \caption{Posterior of $\log_{10}k_0$ for an injection of only $+/\times$ polarisations into the detectors. It is consistent with the lower bound of the $k_0$ prior, indicating there is no signal in the null stream.}
    \label{fig:log10k0_posterior_GR_only}
\end{figure}

In order to illustrate that our analysis is robust to false positives (an analysis that is robust to false positives is sometimes referred to as ``safe''), we repeat the injection study, but with only a $+,\times$-polarised source.
Since the signal is allowed by GR, we expect it to be absent from the null stream. Figure \ref{fig:log10k0_posterior_GR_only} shows that the posterior for $k_0$ is consistent with the lower bound of the $k_0$ prior and hence no signal in the null stream.

\section{Analysis of GWTC-3}\label{sec:data}
We analyse 42 three-detector events in the third gravitational-wave transient catalogue (GWTC-3) \citep{GWTC-3_paper}.
These 42 events are chosen by taking all three-detector events in GWTC-3 with a ringdown frequency less than $\unit[512]{Hz}$. 

We perform parameter estimation on the GW events to obtain the posteriors of the extrinsic parameters of the binary source using the \imrphenompv waveform \citep{PhenomP_paper, PhenomP_paper2}. We use a uniform prior between $[0.05, 1]$ for our mass ratio, a uniform prior between $\unit[[4.6, 250]]{M_\odot}$ for our chirp mass, a uniform prior between $[0, 0.99]$ for our spin magnitudes and a power law prior with $\alpha=2$ between $\unit[[10, 30000]]{\text{Mpc}}$ for luminosity distance. We use the default priors from the parameter estimation runs of GWTC-2 paper \citep{GWTC-2_paper} for the rest of the GW parameters: right ascension, declination, polarisation angle, time of coalescence, phase, inclination angle, tilt angles and azimuthal angles. We marginalise over time, phase and distance during the sampling. We construct the null stream using the maximum likelihood estimate of the extrinsic parameters. We choose the same priors for the Gaussian process hyper-parameters as Section \ref{sec:simulations}. 
The GW strain data used in the parameter estimation is publicly available on the Gravitational Wave Open Science Centre~\citep[GWOSC;][]{open_data_o1_o2, open_data_03}. 

\begin{figure}
    \centering
    \includegraphics[width=0.95\linewidth]{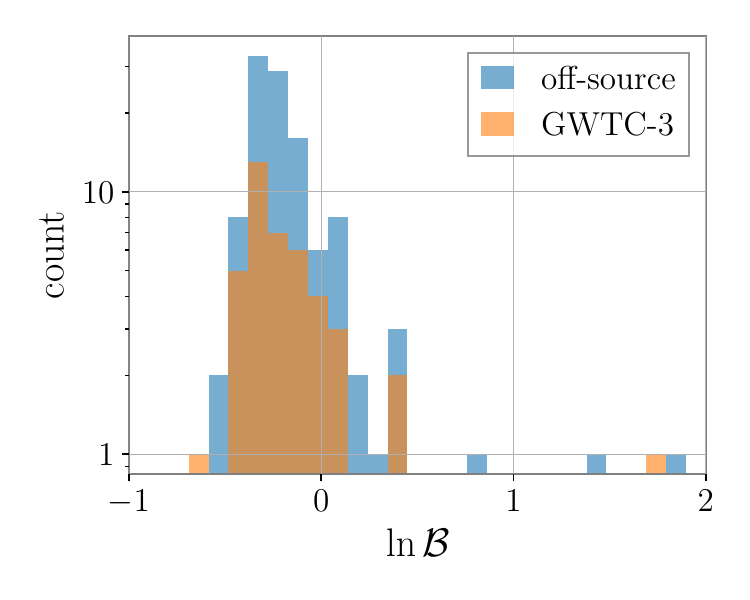}
    \caption{Natural log Bayes factors of 112 off-source segments (blue) and 42 three-detector events in GWTC-3 (orange). All GWTC-3 events have a small Bayes factor with the highest of $\ln\mathcal{B}=1.79$, indicating no signal. The background study show multiple off-source data segments exhibit larger Bayes factors than the GWTC-3 events due to non-stationary noise.}
    \label{fig:hist_BF}
\end{figure}

We obtain the Bayes factor for each of the  42 events (shown in Figure \ref{fig:hist_BF} in blue). 
All the events in GWTC-3 have modest natural log Bayes factors $\lesssim 2$. This implies that there are no detectable signals in the null stream in GWTC-3, which is consistent with the results of previous searches for extra polarisations using null streams in GWTC-2 \citep{TGR_GWTC2} and GWTC-3 \citep{TGR_GWTC3}. The Bayes factors for each individual event is listed in Table \ref{tab:combined_two_columns} in the Appendix.

We place constraints on the strain of a non-GR signal in the null stream using the following method:
\begin{enumerate}
    \item We create 100 different kernels drawn from the kernel parameter posteriors for an event and make a draw from each kernel to reconstruct a time domain strain. We calculate the upper limit of the maximum strain $\delta s_{\text{max}}$ at 90\% credibility. 
    \item We make 100 draws from the posteriors of GW parameters for an event and construct a time domain strain.  We calculate the upper limit of the maximum strain $h_{\text{max}}$ at 90\% credibility. 
    \item We calculate the fractional deviation of the GW strain as $\delta s_{\text{max}}/h_{\text{max}}$.
\end{enumerate}

We calculate $\delta s_{\text{max}}/h_{\text{max}}$ for all three-detector GWTC-3 events, which we list in Table \ref{tab:combined_two_columns} in the Appendix. We find that the highest fractional deviation is $\delta s_{\text{max}}/h_{\text{max}}=7.90$, and the lowest fractional deviation is $\delta s_{\text{max}}/h_{\text{max}}=0.39$.

The lowest fractional deviation is higher than the fractional deviation reported in \citet{Passenger_2025}: $\delta s_{\text{max}}/h_{\text{max}}=0.07$. These different values reflect significant differences in what each analysis seeks to accomplish.
This analysis, which is designed to detect non-standard polarisation modes, requires three observatories to construct a null stream, and so the sensitivity is limited by the least sensitive observatory.
The \cite{Passenger_2025} analysis, on the other hand, is designed to search for any deviation from our compact binary templates.
The sensitivity is always improved by the addition of more observatories.

To see how our results for the three-detector events in GWTC-3 compare with the background noise, we analyse 112 segments of off-source data near GWTC-3 events. We specifically choose $\unit[100]{s}$ and $\unit[200]{s}$ before each events' time of coalescence and $\unit[100]{s}$ after for each event, while ensuring the data segments pass the category-1 (CAT1) and category-2 (CAT2) data quality checks used in LVK's third observing run \citep{LV_detector_noise,Davis_2021,GWTC-3_paper}. We also check they are not overlapping with a glitch recorded in the GravitySpy glitch database \citep{Zevin_2017, Glanzer_2023}. We ensure that there is no glitch within $\unit[1.5]{s}$ of the center of each data segment.

To ensure the analysis of the noise and the GWTC-3 events are as similar as possible, we inject a GW150914-like gravitational-wave signal into each noise segment. We perform parameter estimation with identical priors and settings as our GWTC-3 parameter estimation. 

Figure \ref{fig:hist_BF} shows a histogram of $\ln \mathcal{B}$ for the 112 off-source data and the 42 three-detector events. 
We check if the two distributions are consistent using a Kolmogorov-Smirnov test, and determine that they are, indeed, consistent ($p=0.93$).

\section{Conclusion}\label{sec:conclusion}
The null stream provides a powerful tool to test for deviations from GR.
If non-standard polarisations exist in nature, they can be detected in the null stream without risk of false positive detections from waveform misspecification.
We employ the Gaussian process framework from \citet{Passenger_2025} to search for extra polarisations, enforcing our prior beliefs that any deviations from GR are likely to be localised in time near the merger with a characteristic frequency similar to the merger frequency.
We find no evidence for deviations from GR in GWTC-3 and set limits on the strain amplitude of such deviations.

Data from at least three observatories is required to construct a null stream.
The sensitivity of our search is limited by the noise curve of the third-most sensitive observatory.
Previous work has highlighted the benefits of a sensitive, three-observatory network for sky localisation. 
Such networks are also valuable to search for non-standard polarisations.

\section*{Acknowledgement}
This is LIGO document DCC P2500391.
We acknowledge support from the Australian Research
Council (ARC) Centres of Excellence CE170100004 and CE230100016, as well as ARC
LE210100002, and ARC DP230103088. S.Y.C. and L.P.
receives support from the Australian Government Research
Training Program. This material is
based upon work supported by NSF’s LIGO Laboratory
which is a major facility fully funded by the National
Science Foundation. The authors are grateful for computational resources provided by the LIGO Laboratory
and supported by National Science Foundation Grants
PHY-0757058 and PHY-0823459.

This research has made use of data or software obtained from the Gravitational Wave Open Science Center (gw-openscience.org), a service of LIGO Laboratory,
the LIGO Scientific Collaboration, the Virgo Collaboration, and KAGRA. LIGO Laboratory and Advanced
LIGO are funded by the United States National Science Foundation (NSF) as well as the Science and Technology Facilities Council (STFC) of the United Kingdom, the Max-Planck-Society (MPS), and the State of
Niedersachsen/Germany for support of the construction
of Advanced LIGO and construction and operation of
the GEO600 detector. Additional support for Advanced
LIGO was provided by the Australian Research Council.
Virgo is funded, through the European Gravitational
Observatory (EGO), by the French Centre National
de Recherche Scientifique (CNRS), the Italian Istituto
Nazionale di Fisica Nucleare (INFN) and the Dutch
Nikhef, with contributions by institutions from Belgium,
Germany, Greece, Hungary, Ireland, Japan, Monaco,
Poland, Portugal, Spain. The construction and operation of KAGRA are funded by Ministry of Education,
Culture, Sports, Science and Technology (MEXT), and
Japan Society for the Promotion of Science (JSPS), National Research Foundation (NRF) and Ministry of Science and ICT (MSIT) in Korea, Academia Sinica (AS)
and the Ministry of Science and Technology (MoST) in
Taiwan.

\bibliographystyle{aasjournal}
\bibliography{refs}

\appendix

\section{Marginalising over uncertainty in extrinsic parameters}\label{appendix_A}

The construction of a null stream with a three-detector network depends on four extrinsic parameters: the right ascension $\alpha$, declination $\delta$, polarisation angle $\psi$ and time of coalescence $t_c$. Hence, the likelihood for the Gaussian process is
\begin{align}
    {\cal L}(d_{\text{null}} |\theta, \Lambda) = &
    \frac{1}{2\pi \det \textbf{\textit{C}}(\Lambda)} \exp 
    \left(- \left| \frac{1}{2} d_\text{null}^\dagger (\theta)\textbf{\textit{C}}^{-1} (\theta, \Lambda) d_{\text{null}}(\theta) \right| \right),
\end{align}
where $\theta = \{\alpha, \delta, \psi, t_c \}$.

The covariance matrix $C(\theta, \Lambda)$ depends on $\theta$ 
\begin{align}\label{eq:C2}
    \textbf{\textit{C}}(\theta, \Lambda) = \textbf{\textit{S}}(\theta, \Lambda) + \textbf{\textit{N}}(\theta),
\end{align}
as the signal matrix is dependent on $\theta$
\begin{align}\label{eq:Sij_theta}
    \textbf{\textit{S}}_{ij}(\theta, \Lambda) = & 
    \left( \sum_{m} (F_m^1(\theta) - \eta(\theta)F_m^2(\theta) - \zeta(\theta) F_m^3(\theta))^2 
    \right) \textbf{\textit{K}}_{ij}(\Lambda),
\end{align}
and the noise matrix $ \textbf{\textit{N}}_{ij}(\theta)$ is now a function of the $\theta$ as the null stream power spectral density is dependent on $\theta$ (see Eq. \ref{eq:PSD}). We can marginalise over the four parameters which accounts for the uncertainty in the null stream construction
\begin{align}\label{eq:L_sky_marg}
    {\cal L}(d_\text{null} |\Lambda) = \int {\cal L}(d_\text{null} |\theta, \Lambda) p(\theta) d\theta.
\end{align}

The integral can be approximated as a sum over samples $\theta$ for each kernel parameter $k$
\begin{align}\label{eq:L_sky_marg_discrete}
    {\cal L}(d_\text{null} |\Lambda_k) \approx \frac{1}{N_{\theta}}\sum^{N_\theta}_i {\cal L}(d_\text{null} |\theta_i, \Lambda_k),
\end{align}
where $N_{\theta}$ is the number of samples for $\theta$. We can use the kernel parameter (proposal) posterior from the maximum likelihood estimate of the extrinsic parameters $\theta_{\text{max}}$ to approximate the (target) posterior marginalised over the extrinsic parameter. We use a method called reweighting which involves calculating weights,
\begin{align}\label{eq:weights}
    w(d|\Lambda_k)=\frac{1}{N_{\theta}}\sum^{N_\theta}_i\frac{{\cal L}(d_\text{null} |\theta_i,\Lambda_k)}{{\cal L}(d_\text{null} |\theta_{\text{max}}, \Lambda_k)},
\end{align}
which we use to reweight the proposal posterior to the target posterior
\begin{align}
    p(\Lambda_k| d_{\text{null}})= w(d|\Lambda_k) p(\Lambda_k|d_{\text{null}}, \theta_{\text{max}}).
\end{align}
When performing this method, one should ensure that the number of effective samples is sufficiently large \citep{Payne_2019}. 

\begin{figure*}[t]
    \centering
    \includegraphics[width=0.7\linewidth]{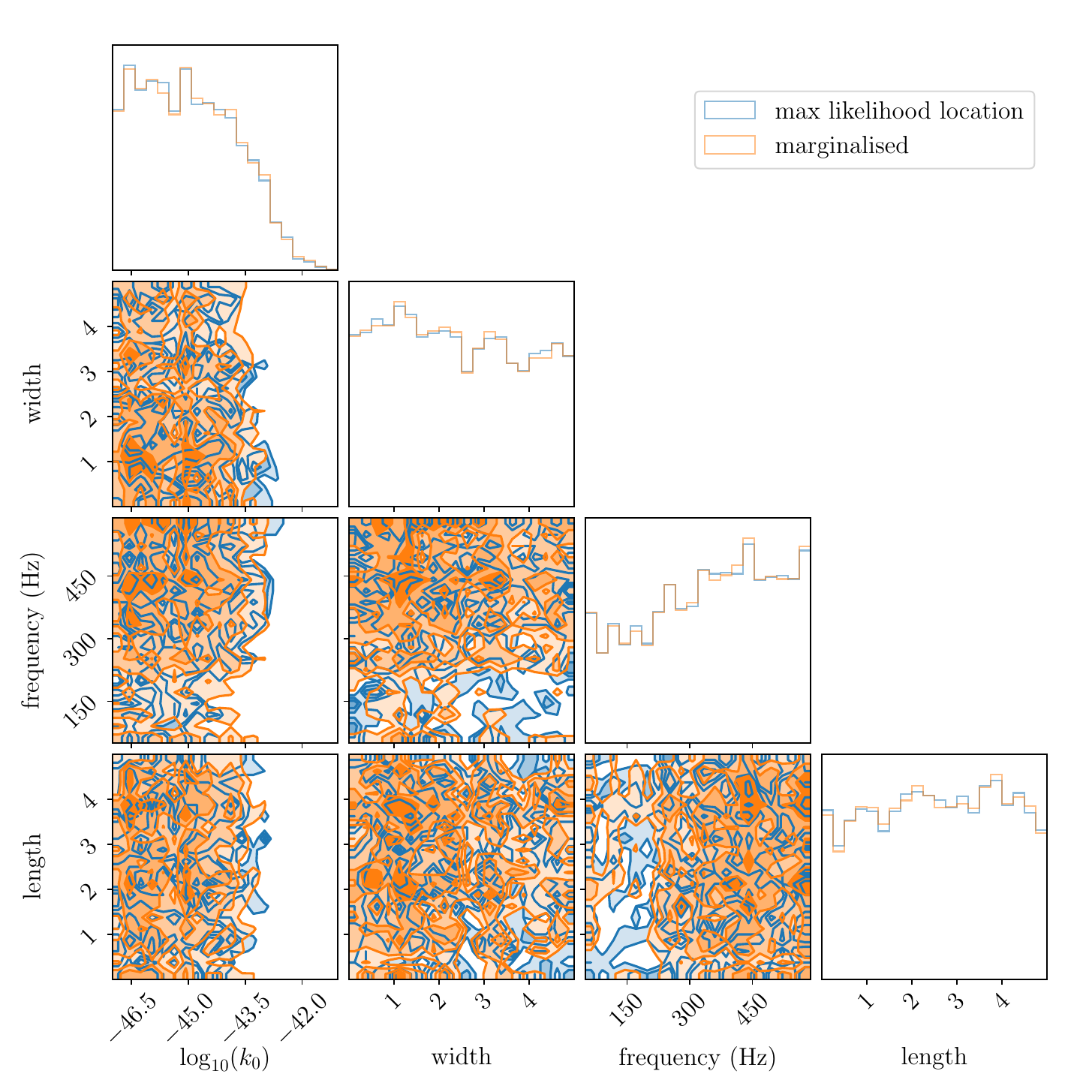}
    \caption{Posterior distributions for the kernel parameters describing the event GW170814. 
    In blue we show the result obtained by calculating the null stream with the maximum-likelihood parameters for right ascension, declination, polarisation angle and time of coalescence.
    In orange we show the result obtained by marginalising over uncertainty in the these parameters. 
    The similarity between orange and blue shows that uncertainty in sky location has a marginal effect on the posterior distribution for the kernel parameters. }
    \label{fig:corner_compare_sky_loc}
\end{figure*}

We investigate the effect of different sky locations, polarisations angles and times of coalescence on our results. We choose the event GW170814 and run our analysis with two different methods: taking the maximum likelihood values from the parameter estimation results and averaging the posterior distribution of the kernel parameters over 10 randomly drawn samples. The former method does not account for the uncertainty in the extrinsic parameters while the latter is an approximation to marginalising over the kernel parameters. Figure \ref{fig:corner_compare_sky_loc} compares the posterior distributions of the kernel parameters. The two results are consistent with one another. Hence, we conclude that the effect of the potential uncertainty in the sky location, polarisation angle and time of coalescence due to the binary-merger analysis on our results is negligible and marginalisation is not needed.

\section{Bayes factors and fractional deviation for GWTC-3}
Table \ref{tab:combined_two_columns} is a summary of the natural log Bayes factors $\ln \mathcal{B}$ of 42 three-detector events in GWTC-3. We find that the $\ln \mathcal{B}$ of 42 events are small ($\ln\mathcal{B}<8$), indicating no evidence of extra polarisations. To constrain the amplitude of a potential deviation of GR, we calculate $\delta s_{\text{max}}/h_{\text{max}}$  for each three-detector event in GWTC-3. This is shown in Table \ref{tab:combined_two_columns}. We find that the lowest upper limit is for the event GW190602\_175927 with $\delta s_{\text{max}}/h_{\text{max}}<0.39$.

\begin{table}[h]
    \centering
    \begin{minipage}{0.48\textwidth}
        \centering
        \begin{tabular}{|c|c|c|}
            \hline
            Event & $\ln \mathcal{B}$ & $\delta s_{\text{max}}/h_{\text{max}}$ \\
            \hline
            GW170729 & 0.08 & 2.08 \\
            GW170809 & -0.12 & 0.48 \\
            GW170814 & -0.33 & 0.63 \\
            GW170818 & -0.21 & 0.81 \\
            GW190408\_181802 & -0.28 & 1.71 \\
            GW190412\_053044 & 0.03 & 2.27 \\
            GW190413\_134308 & 0.06 & 3.07 \\
            GW190503\_185404 & -0.16 & 0.57 \\
            GW190512\_180714 & -0.20 & 3.59 \\
            GW190513\_205428 & 1.79 & 3.74 \\
            GW190517\_055101 & -0.45 & 0.43 \\
            GW190519\_153544 & -0.42 & 0.79 \\
            GW190521\_030229 & -0.11 & 1.46 \\
            GW190602\_175927 & -0.41 & 0.39 \\
            GW190701\_203306 & -0.40 & 0.41 \\
            GW190706\_222641 & -0.07 & 1.24 \\
            GW190727\_060333 & -0.20 & 1.31 \\
            GW190803\_022701 & -0.28 & 1.03 \\
            GW190828\_063405 & 0.11 & 3.79 \\
            GW190828\_065509 & -0.28 & 3.17 \\
            GW190915\_235702 & -0.24 & 0.71 \\
            \hline
        \end{tabular}
    \end{minipage}%
    \hfill
    \begin{minipage}{0.48\textwidth}
        \centering
        \begin{tabular}{|c|c|c|}
            \hline
            Event & $\ln \mathcal{B}$ & $\delta s_{\text{max}}/h_{\text{max}}$ \\
            \hline
            GW190916\_200658 & 0.36 & 5.29 \\
            GW190926\_050336 & -0.31 & 1.95 \\
            GW190929\_012149 & -0.24 & 1.97 \\
            GW191113\_071753 & -0.13 & 1.92 \\
            GW191127\_050227 & -0.28 & 0.77 \\
            GW191215\_223052 & -0.34 & 1.37 \\
            GW191219\_163120 & -0.61 & 2.77 \\
            GW191230\_180458 & -0.15 & 2.08 \\
            GW200129\_065458 & 0.35 & 0.47 \\
            GW200208\_130117 & -0.30 & 1.24 \\
            GW200208\_222617 & -0.23 & 1.55 \\
            GW200209\_085452 & -0.30 & 1.24 \\
            GW200210\_092254 & -0.34 & 7.90 \\
            GW200216\_220804 & -0.39 & 0.59 \\
            GW200219\_094415 & -0.30 & 1.69 \\
            GW200220\_061928 & -0.35 & 1.72 \\
            GW200224\_222234 & -0.11 & 0.77 \\
            GW200308\_173609 & -0.01 & 6.39 \\
            GW200311\_115853 & -0.37 & 0.54 \\
            GW200316\_215756 & -0.30 & 4.79 \\
            GW200322\_091133 & -0.04 & 6.07 \\
            \hline
        \end{tabular}
    \end{minipage}
    \caption{Natural-log Bayes factors and fractional deviations $\delta s_{\text{max}}/h_{\text{max}}$ for selected GWTC-3 events.}
    \label{tab:combined_two_columns}
\end{table}

\end{document}